\documentclass[a4paper]{spie}  

\usepackage[]{graphicx}
\usepackage{amssymb,amsmath,psfrag,url}

\newcommand{\Field}[1]{{\boldsymbol{#1}}}
\newcommand{\Kvec}[1]{{\vec{#1}}}

\title{Rigorous Simulation of 3D Masks} 

\author{
Sven Burger\supit{\,ab}, 
Roderick K\"ohle\supit{\,c}, 
Lin Zschiedrich\supit{\,ab},
Hoa Nguyen\supit{\,d},\\
Frank Schmidt\supit{\,ab}, 
Reinhard M\"arz\supit{\,be}, 
Christoph N\"olscher\supit{\,d}
\skiplinehalf
\supit{a}
Zuse Institute Berlin,
Takustra{\ss}e 7,
D\,--\,14\,195 Berlin,
Germany\\
DFG Forschungszentrum {\sc Matheon},
Stra{\ss}e des 17.\,Juni 136, 
D\,--\,10\,623 Berlin,
Germany
\smallskip\\
\supit{b}
JCMwave GmbH,
Haarer Stra{\ss}e 14a,
D\,--\,85\,640 Putzbrunn, 
Germany
\smallskip\\
\supit{c}
Qimonda AG,
Advanced Technology Software\\
Am Campeon 1-12, D\,--\,85\,579 M\"unchen, Germany
\smallskip\\
\supit{d}
Qimonda Dresden GmbH \& Co.OHG, QD P LM F\\
K\"onigsbr\"ucker Stra{\ss}e 180,
D\,--\,01\,099 Dresden,
Germany
\smallskip\\
\supit{e}
Infineon Technologies AG,
COM CAL D TD RETM PI,\\
Balanstra{\ss}e 73,
D\,--\,81\,541 M\"unchen,
Germany
}

\authorinfo{
Corresponding author: S. Burger\\
URLs: http://www.zib.de/Numerik/NanoOptics\\
http://www.jcmwave.com\\
Email: burger@zib.de
}

\begin{document} 
  \maketitle 
\noindent
Copyright 2006  Society of Photo-Optical Instrumentation Engineers.\\
This paper has been published in Proc.~SPIE {\bf 6349}, 63494Z
(2006),  
({\it 26th Annual BACUS Symposium on Photomask Technology, 
P.~M.~Martin, R. J. Naber, Eds.})
and is made available 
as an electronic reprint with permission of SPIE. 
One print or electronic copy may be made for personal use only. 
Systematic or multiple reproduction, distribution to multiple 
locations via electronic or other means, duplication of any 
material in this paper for a fee or for commercial purposes, 
or modification of the content of the paper are prohibited.

\begin{abstract}

We perform 3D lithography simulations by using a finite-element 
solver.
To proof applicability to real 3D problems we investigate
DUV light propagation through a structure of size 9\,$\mu$m $\times$
 4\,$\mu$m $\times$  65\,nm.
On this relatively large computational domain we 
perform rigorous computations (No Hopkins) taking into account 
a grid of $11\times 21$ source points with two polarization directions 
each. 
We obtain well converged results with an accuracy of the 
diffraction orders of about~1\%.
The results compare  well to experimental aerial imaging results.
We further investigate the convergence of 3D solutions towards 
quasi-exact results obtained with different methods.

\end{abstract}

\keywords{Photomask, microlithography, simulation, finite element method, FEM}

\section{Introduction}
Shrinking feature sizes in optical lithography lead to increasing 
importance of rigorous simulations for process design~\cite{Erdmann2004a}. 
Modern lithography simulators include modules describing
illumination, transfer of the optical field through the mask and 
aberrating optical system of the lithographic equipment, the propagation inside
the photoresist, the processes leading to the resist image
and -- in advanced systems -- the etching processes 
leading to the etched image. 
After nearly two decades
of lithography simulation, most of the modules along the simulation 
chain have attained a high degree of maturity.
However, the simulation of light propagation through lithography masks
is still challenging in terms of computational time and memory 
and accuracy of the results.

The computation of the print image of a whole chip remains extremely demanding
although approximations, multi-threading and even hardware accelerators 
are applied to reduce the runtime of simulations. 
Rigorous simulations are restricted today to small areas
and even those simulations suffer from the high computational effort. 
At the same time, the progress on the semiconductor roadmap forces the need of 
rigorous 3D simulations.

Keeping this background in mind, we employed a frequency-domain 
finite-element method (FEM) solver for Maxwell's equations.
In a recent benchmark this solver has been shown to be superior in accuracy and 
computational time requirements by several 
orders of magnitude, compared to a FDTD solver~\cite{Burger2005bacus}. 
Further, this solver
has been successfully applied to a wide range of 3D
electromagnetic field computations including
left-handed metamaterials in the optical regime~\cite{Enkrich2005a,Burger2005w},
photonic crystals~\cite{Burger2005a}, and
nearfield-microscopy~\cite{Kalkbrenner2005a}.

\begin{figure}[htb]
\centering
\psfrag{w}{\sffamily w}
\psfrag{d}{\sffamily d}
\psfrag{l_1}{\sffamily $\mbox{l}_1$}
\psfrag{l_2}{\sffamily $\mbox{l}_2$}
\psfrag{l_3}{\sffamily $\mbox{l}_3$}
\psfrag{p_x}{\sffamily $\mbox{p}_x$}
\psfrag{p_y}{\sffamily $\mbox{p}_y$}
\includegraphics[width=0.3\textwidth]{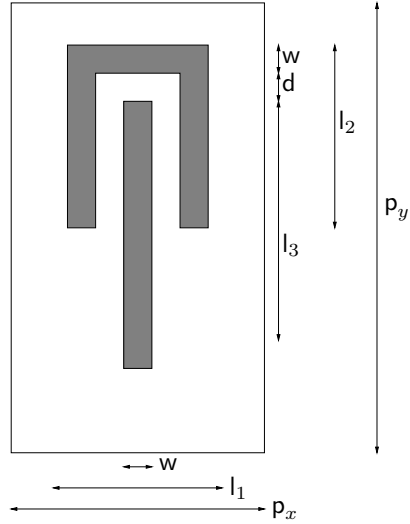}
\caption{
Schematics of the 3D test structure. 
The height of the lines is 65.4\,nm, the lateral size of the 
computational window is $p_x\times p_y = 4\,\mu\mbox{m}\times 9\,\mu\mbox{m}$. 
}
\label{schema_fork}
\end{figure}

We consider light scattering  off a mask which is 
periodic in the $x-$ and $y-$directions and is enclosed by  homogeneous 
substrate (at $z_{sub}$) and superstrate (at $z_{sup}$) 
which are infinite in the $-$, resp.~$+z-$direction. 
However, the presented FEM concept holds as well for non-periodic scattering 
objects, where the surrounding space is either homogeneous or consists 
of layered media or waveguide structures. 
Light propagation in the investigated system is governed by Maxwell's equations
where  vanishing densities of free charges and currents are assumed. 
The dielectric coefficient $\varepsilon(\vec{x})$ and the permeability 
$\mu(\vec{x})$ of the considered photomasks are periodic and complex, 
$\varepsilon \left(\vec{x}\right)  =  \varepsilon \left(\vec{x}+\vec{a} \right)$, 
$\mu \left(\vec{x} \right)  =  \mu \left(\vec{x}+\vec{a} \right)$.
Here $\vec{a}$ is any elementary vector of the periodic lattice.  
For given primitive lattice vectors 
$\vec{a}_{1}$ and $\vec{a}_{2}$ an elementary cell 
$\Omega\subset\mathbb R^{3}$ is defined as
$\Omega = \left\{\vec{x} \in \mathbb R^{2}\,|\,
x=\alpha_{1}\vec{a}_1+\alpha_{2}\vec{a}_2;
0\leq\alpha_{1},\alpha_{2}<1
\right\}
\times [z_{sub},z_{sup}].$
A time-harmonic ansatz with frequency $\omega$ and magnetic field 
$\Field{H}(\vec{x},t)=e^{-i\omega t}\Field{H}(\vec{x})$ leads to
the following equations for $\Field{H}(\vec{x})$:
\begin{itemize}
\item
The wave equation and the divergence condition for the magnetic field:
\begin{eqnarray}
\label{waveequationH}
\nabla\times\frac{1}{\varepsilon(\vec{x})}\,\nabla\times\Field{H}(\vec{x})
- \omega^2 \mu(\vec{x})\Field{H}(\vec{x}) &=& 0,
\qquad\vec{x}\in\Omega,\\
\label{divconditionH}
\nabla\cdot\mu(\vec{x})\Field{H}(\vec{x}) &=& 0,
\qquad\vec{x}\in\Omega .
\end{eqnarray}
\item
Transparent boundary conditions at the boundaries to the 
substrate (at $z_{sub}$) and superstrate (at $z_{sup}$), $\partial\Omega$,
where $\Field{H}^{in}$ is the incident magnetic field (plane wave 
in this case), and $\vec{n}$ is the normal vector on $\partial\Omega$:
\begin{equation}
\label{tbcH}
	\left(
        \frac{1}{\varepsilon(\vec{x})}\nabla \times (\Field{H} - 
        \Field{H}^{in})
	\right)
	\times \vec{n} = DtN(\Field{H} - 
        \Field{H}^{in}), \qquad \vec{x}\in \partial\Omega.
\end{equation}
The $DtN$ operator (Dirichlet-to-Neumann) is  realized with 
the PML method~\cite{Zschiedrich03}. 
This is a generalized formulation of Sommerfeld's radiation condition; it
can be realized alternatively by the Pole condition method~\cite{Hohage03a}.
\item
Periodic boundary conditions for the transverse boundaries, $\partial\Omega$,
governed by Bloch's theorem~\cite{Sakoda2001a}:
\begin{equation}
\label{bloch}
\Field{H}(\vec{x}) = e^{i \Kvec{k}\cdot\vec{x}} \Field{u}(\vec{x}), \qquad
\Field{u}(\vec{x})=\Field{u}(\vec{x}+\vec{a}),
\end{equation}
where the Bloch wavevector $\Kvec{k}\in\mathbb{R}^3$ is defined by the
incoming plane wave $\Field{H}^{in}$.

\end{itemize}
\begin{figure}[t]
(a)\includegraphics[width=0.31\textwidth]{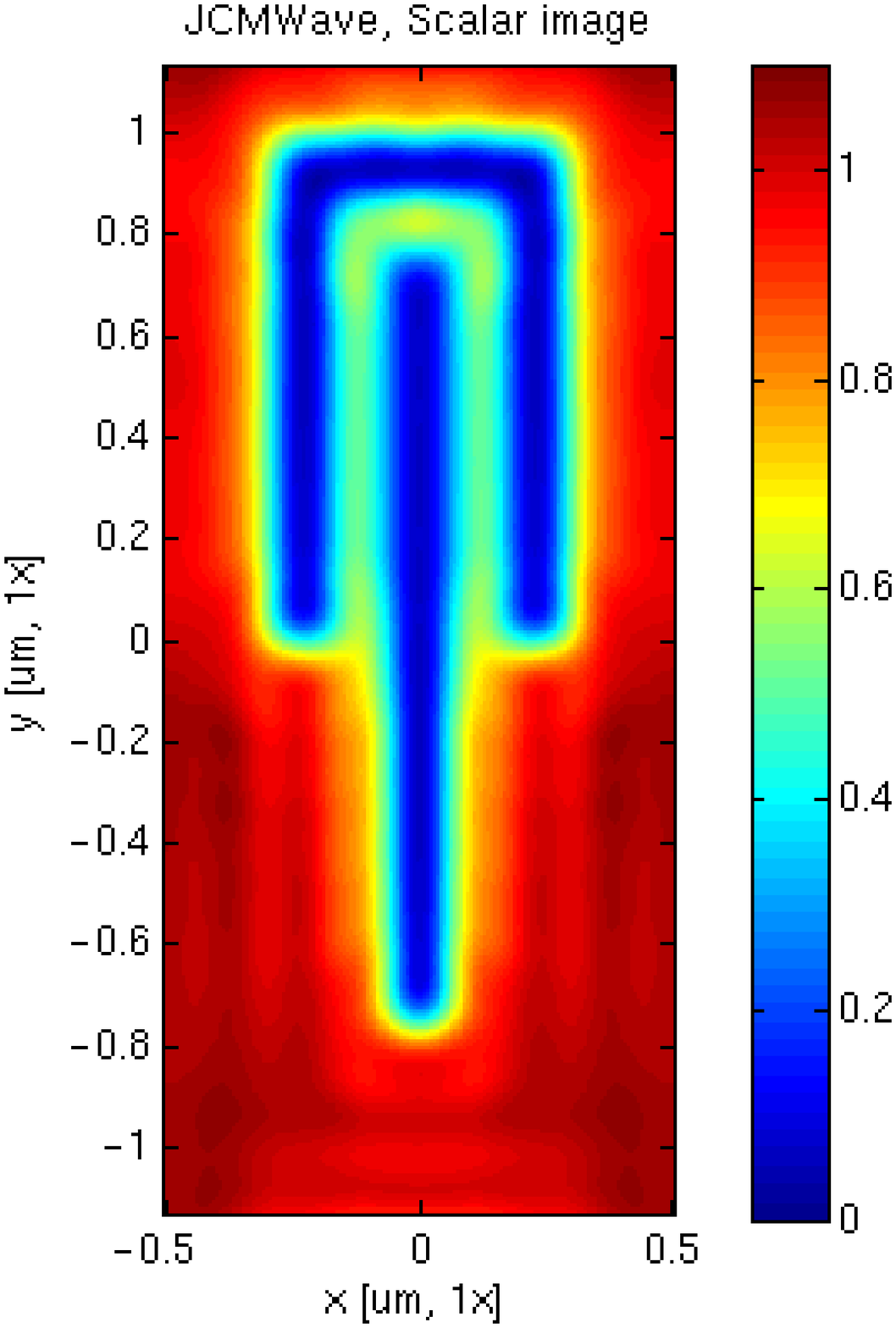}
(b)\includegraphics[width=0.31\textwidth]{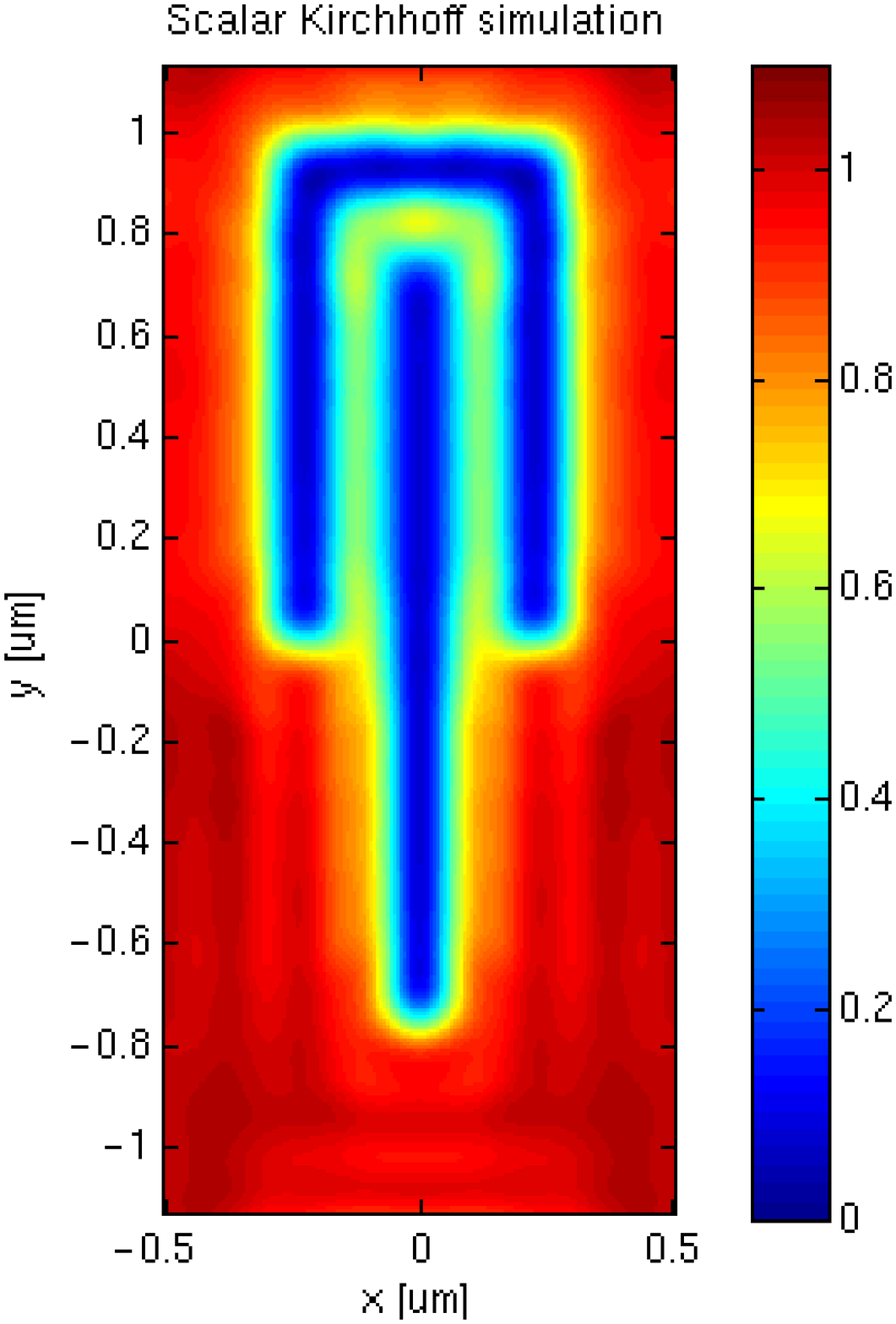}
(c)\includegraphics[width=0.31\textwidth]{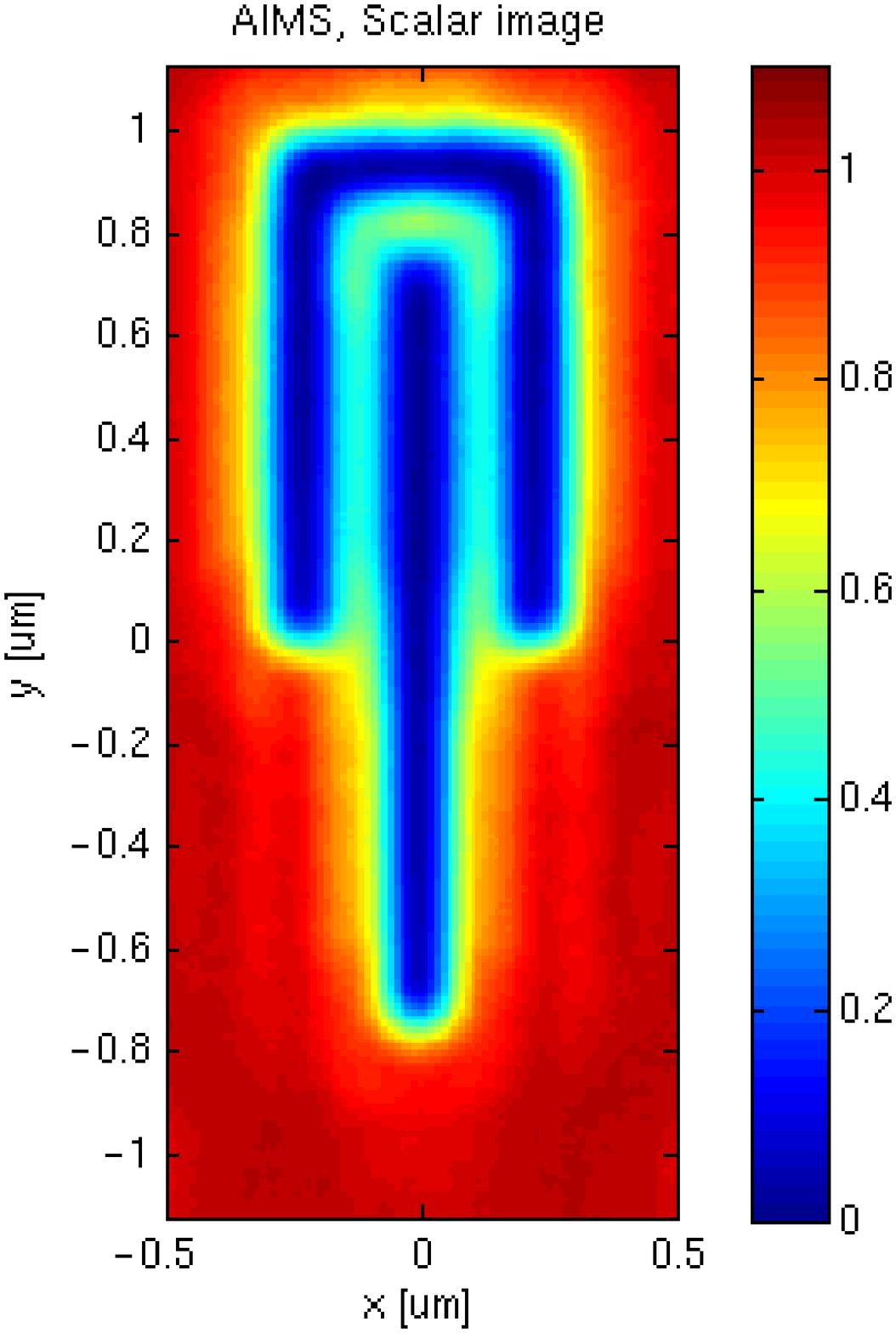}
\caption{
Aerial image of the test-structure depicted in Fig.~\ref{schema_fork}.
Rigorous FEM simulation (JCMharmony) (a), simulation using thin mask approximation (Kirchhoff) (b), 
and experimentally attained image (AIMS) (c) agree well.
}
\label{aerial_image}
\end{figure}

Similar equations are found for the electric field 
$\Field{E}(\vec{x},t)=e^{-i\omega t}\Field{E}(\vec{x})$;
these are treated accordingly.
The finite-element method solves Eqs.~(\ref{waveequationH}) -- (\ref{bloch})
in their weak form, i.e., in an integral representation. 
The computational domain is discretized with triangular (2D)
or tetrahedral/prismatoidal (3D) patches. 
The use of prismatoidal patches is well suited for layered geometries, as in 
photomask simulations. 
This also simplifies the geometry description of the mask layout. 
Sidewall angles different from 90\,deg are not regarded throughout this paper;
however, they can easily be implemented with reasonable restrictions. 
The function spaces are discretized using Nedelec's edge elements, 
which are vectorial functions of polynomial order (here, first to fourth order) defined 
on the triangular or tetrahedral patches~\cite{Monk2003a}. 
In a nutshell, FEM can be explained as expanding the field 
corresponding to the exact solution of Equation~(\ref{waveequationH}) in the 
basis given by these elements.
This leads to  a large sparse matrix equation (algebraic problem).
For details on the weak formulation, 
the choice of Bloch-periodic functional spaces,
the FEM discretization, and our implementation of the PML
method we refer to previous works~\cite{Zschiedrich03,Burger2005a,Zschiedrich2005a}.
In future implementations performance will further be increased 
by using higher order elements, $p>4$, 
$hp$-adaptive methods, and by using elements of different polynomial order parallel 
and orthogonal to the layers of the layered geometry (corresponding to the 
$x-y-$plane of Fig.~\ref{schema_2d3d}\,b). 
To solve the algebraic problem on a standard workstation 
either standard linear algebra decomposition techniques (LU-factorization, e.g.,
package PARDISO~\cite{PARDISO})
or iterative and domain decomposition 
methods~\cite{Deuflhard2003a,Zschiedrich2005b,Zschiedrich2006b} are used, 
depending on problem size.
Domain decomposition methods as shown in Ref.~\cite{Zschiedrich2005b} for 2D 
FEM simulations can be easily transferred to layered 3D geometries (typical 
photomask geometries) and other 3D geometries.
Due to the use of multi-grid algorithms, the computational time and the memory requirements
grow linearly with the number of unknowns.

\begin{table}[h]
\begin{center}
\begin{tabular}{|l|c|c|}
\hline
parameter & data set 1 & data set 2 \\ 
\hline 
\hline 
$p_x$ & 4000\,nm &  800\,nm \\ 
$p_y$ & 9000\,nm &  800\,nm \\ 
$h$ & 65.4\,nm &    65.4\,nm\\ 
$w$   & 390\,nm &  400\,nm \\ 
$d$   & 520\,nm &  \\ 
$l_1$   & 2210\,nm &  \\ 
$l_2$   & 3910\,nm &  \\ 
$l_3$   & 6000\,nm &  \\ 
\hline 
\hline 
$\lambda_0$ & \multicolumn{2}{c|}{193.0\,nm } \\ 
$n_1$       & \multicolumn{2}{c|}{$2.52+0.596i$} \\ 
$n_2$       & \multicolumn{2}{c|}{$1.56306$} \\ 
\hline 
\end{tabular} 

\caption{Parameter settings for the simulations in 
Section~\ref{section3d} (data set 1) and 
Section~\ref{sectionvalidation} (data set 2). 
}
\label{table_fork}
\end{center}
\end{table}

\section{3D Simulations}

\label{section3d}
\begin{figure}[htb]
\centering
\includegraphics[width=0.5\textwidth]{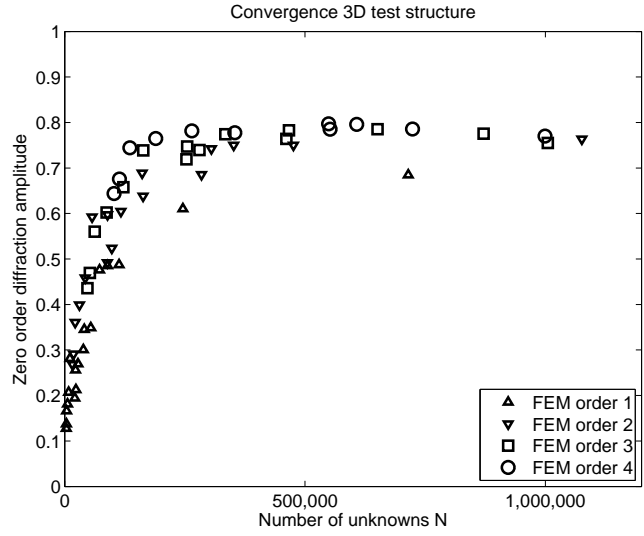}
\caption{ Convergence of the zero order far field coefficient 
for light transition through the test structure described in 
Fig.~\ref{schema_fork} and Table~\ref{table_fork}.
Magnitude of the far field coefficient in dependence on the number of 
ansatz functions for the numerical solution $N$ for finite elements 
of polynomial degree $p=1\dots 4$.
Results of an estimated relative error around 1-2\%
are reached for $N\approx 5\cdot 10^5$ and $p=4$.
}
\label{finger_convergence}
\end{figure}

We investigate a 3D test structure as schematically depicted in Figure~\ref{schema_fork}.
The structure consists of MoSi-lines of height $h$ with a sidewall angle of 90\,deg 
on a glass substrate.
This pattern was chosen to have low vector and interference effects but still
 significant 3D effects, and because it appears in current  lithography production. 
With the  project parameters as given in Table~\ref{table_fork} the size 
of the computational domain is around $10^3$ cubic wavelengths.
We discretize the computational domain using prismatoidal patches, 
and we use higher order, vectorial ansatz functions as finite elements defined on these
patches.

Figure~\ref{finger_convergence} shows the convergence of the simulated zero order 
far field coefficient.
Plotted is the magnitude of the far field coefficient vs.~number of degrees of 
freedom of the finite element expansion. 
Data points are attained using finite elements of first, second, 
third, and fourth polynomial order and using meshes with different refinement levels.
With increasing finite element degree and with decreasing mesh size the results 
converge.
From these results we guess the relative error for a solution using fourth 
order finite elements and using about $5\times 10^5$ unknowns 
is of the order of $1\%$.

To model illumination with a realistic source we construct a grid of 
$11\times 21$ source points in wavevector-space.
For each source point we define two plane waves with orthogonal polarizations. 
This makes a total of 462~sources. 
In order to obtain the scattering response of  an extended, 
 measured C-Quad source we linearly interpolate 
the results for each measured source point between the closest simulated source points
and superpose the results. 
JCMharmony allows to calculate the scattering response of all of the 462 sources in 
a single programme run. 
The total computation time on a workstation (2 64bit processors, around 20~GB RAM)
to obtain the 462 near field solutions, each with 482,040 degrees of freedom, was around 
100~minutes.
We currently work on the significant reduction of the near field computation time. 

A simulation on this area is impossible with present Solid-E~3.3 or Prolith~9.1 
even for a single illumination direction due to memory consumption 
at necessary resolution. Also the simulation time with unacceptable coarse grids is orders 
of magnitude higher.

We use the far field coefficients to generate an aerial image.
Figure~\ref{aerial_image}\,a) shows a pseudo color representation of the 
intensity distribution in the image plane
(demagnification factor 4). 
The scalar intensity distribution has been obtained from the vectorial electric
field distribution.
Figure~\ref{aerial_image}\,b) shows a similar intensity distribution obtained using a 
non-rigorous method (Kirchhoff approximation).
Obviously, the approximation compares well with the rigorous solution for this 
specific simulation example. I.e., interference effects, high-NA effects or other 
effects do not play a role here. 
This example was rather chosen to demonstrate the applicability of a rigoros method 
to large 3D problems. 

Figure~\ref{aerial_image}\,c) shows an experimentally obtained aerial image 
obtained using AIMS.
The minima in the experimental aerial image of the mask structure are more pronounced than in 
the simulation. 
This can be, e.g., caused by uncertainties in the geometry of the sample. 
As has been shown in different works~\cite{Pomplun2006bacus} the high accuracy and 
speed of rigorous FEM simulations can be utilized to obtain precise informations about 
the sample geometry or material parameters by optimizing the deviation from experimentally 
obtained data.

\section{Validation of the Results}
\label{sectionvalidation}
In order to validate the results of the FEM solver we have performed 
several tests. 
Here we present 3D simulations of line masks and compare the results 
to results using 2D simulations of the same physical settings.
For these we have 
we have investigated the convergence towards results 
obtained with various methods, as reported earlier~\cite{Burger2005bacus}. 

\begin{figure}[htb]
\centering
\psfrag{(a)}{\sffamily a)}
\psfrag{(b)}{\sffamily b)}
\psfrag{w}{\sffamily w}
\psfrag{p}{\sffamily $\mbox{p}_x$}
\psfrag{px}{\sffamily $\mbox{p}_x$}
\psfrag{py}{\sffamily $\mbox{p}_y$}
\psfrag{x}{\sffamily x}
\psfrag{y}{\sffamily y}
\psfrag{z}{\sffamily z}
\psfrag{h}{\sffamily h}
\psfrag{b}{} 
\psfrag{air}{\sffamily  air}
\psfrag{substrate}{\sffamily  substrate}
\psfrag{line}{\sffamily  line}
\includegraphics[width=0.5\textwidth]{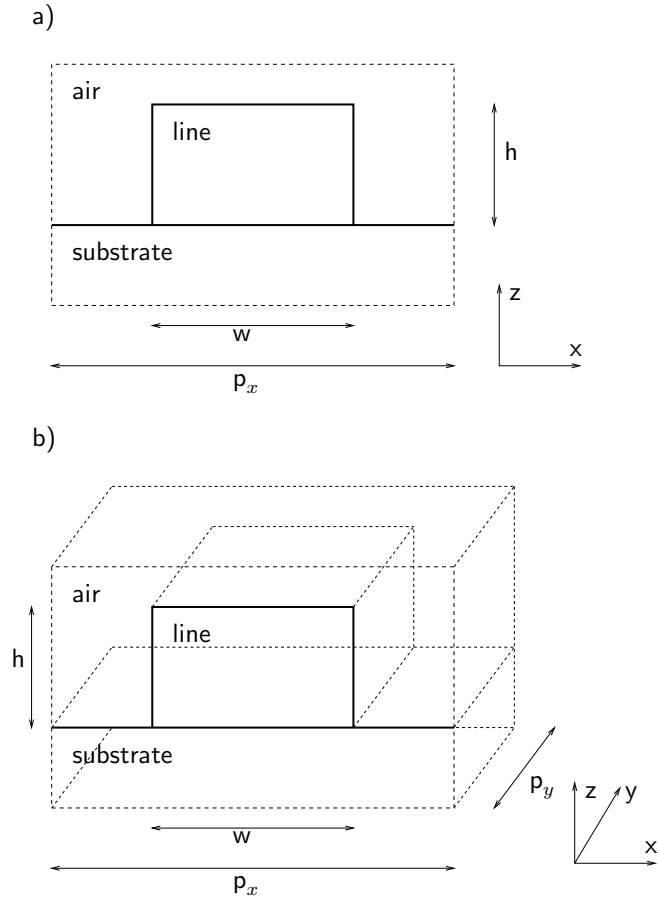}
\caption{
Schematics of the computational domain of a periodic linemask for 
2D calculations (a) and for 3D calculations (b):
The geometry consists of a 
line of width $w$ (at center of the line), height $h$ and  sidewall 
angle $\beta$,  on a  substrate material $SiO_2$, 
surrounded by air.
The geometry is periodic in $x$-direction with a pitch of $p_x$ 
and it is independent on the $y$-coordinate.
The refractive indices of the different present materials are denoted 
by $n_1$ (line), $n_2$ (substrate) and $n_3$ (air), $n_3=1.0$.
}
\label{schema_2d3d}
\end{figure}

\begin{figure}[htb]
\centering
\includegraphics[width=0.5\textwidth]{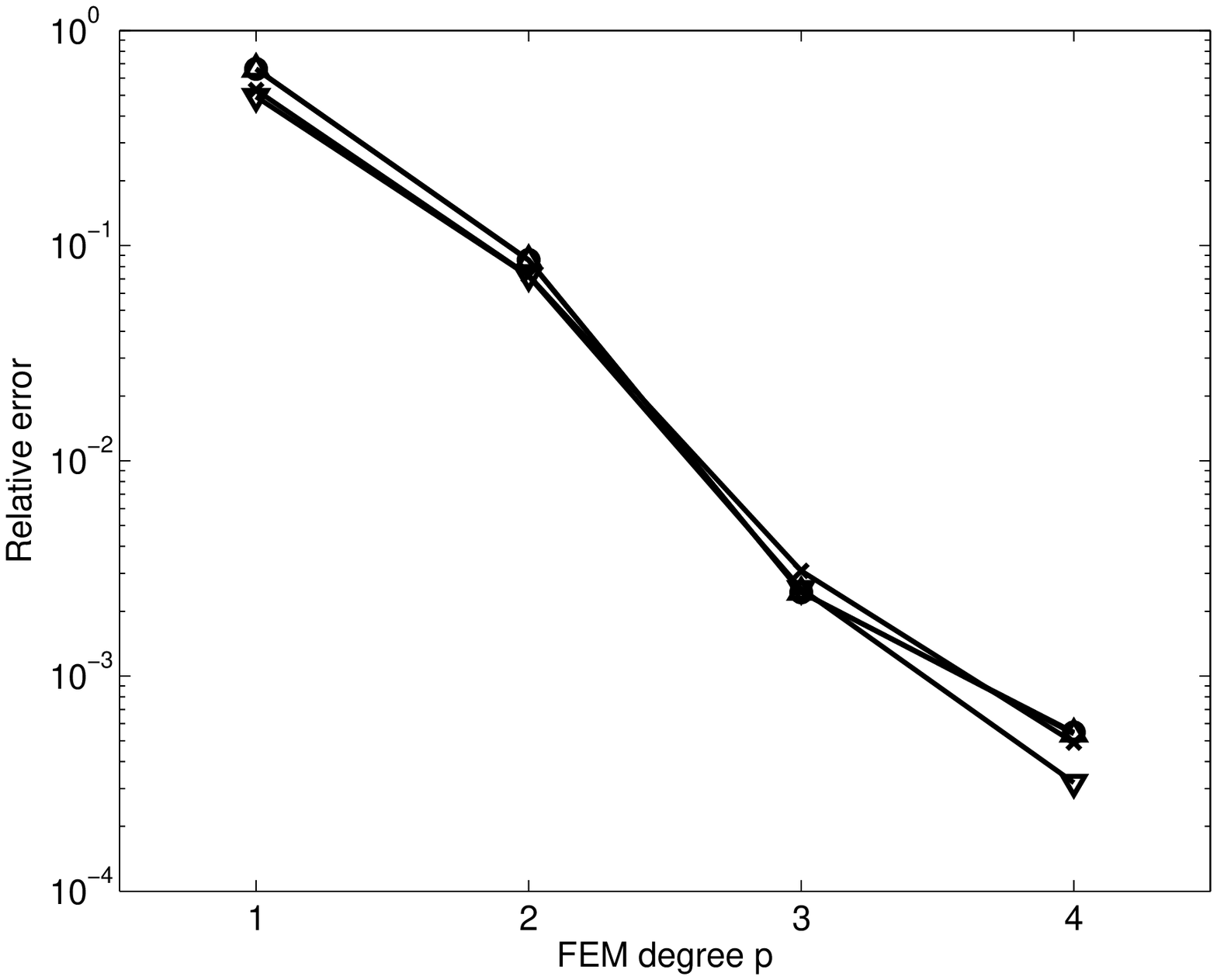}
\caption{ 3D computation of diffraction off a linemask.
Convergence of the results towards the quasi-exact result obtained with 
2D methods. Relative error in dependence of finite element degree $p$.
The different symbols denote triangulations with different mesh sizes 
(typical triangular sidelength $h$ from 30 to 45\,nm).
}
\label{linemask_convergence}
\end{figure}

Figure~\ref{schema_2d3d} shows a schematics of the geometry of a periodic 
line mask. 
TM polarized light is incident from the substrate and is diffracted into various
diffraction orders. 
The geometry parameters are listed in Table~\ref{table_fork} (data set 2).
Please see Reference~\cite{Burger2005bacus} for more details on this test case. 
The 2D simulation results given in the first line of Table~\ref{bm_table_1}
are the best converged results from this reference for the given 
test case. We therefore refer to these results as 
'quasi-exact' results. 
Table~\ref{bm_table_1} further shows results obtained with 3D FEM 
on the geometry as depicted in Figure~\ref{schema_2d3d}\,b).
Vectorial finite elements of order 1 to 4 on grids with different 
mesh refinements have been used to obtain rigorous near field solutions 
from which the zero order far field coefficients are obtained. 
As expected and as can be seen from the results, most accurate results 
are obtained using elements of high order and fine meshes. 
Figure~\ref{linemask_convergence} shows the convergence of the results 
on several fixed FEM grids with elements of 
increasing polynomial order. 
As expected, with increasing polynomial order the numerical 
approximation error converges exponentially towards zero. 

\begin{table}[h]
\begin{center}
\begin{tabular}{|rrlll|}
\hline
\multicolumn{5}{|l|}{JCMharmony 2D (TM)} \\
\hline
 & & $\Re (FC_0)$ & $\Im (FC_0)$ & $|FC_0|$ \\ 
 & &{\bf -0.21943} &{\bf 0.27179} &{\bf 0.34932}  \\ 
\hline \hline
\multicolumn{5} {|l|}{JCMharmony 3D, first order elements (TM)} \\
\hline
n &N & $\Re (FC_0)$ & $\Im (FC_0)$ & $|FC_0|$ \\ 
     1  &  470  &{\bf -0.}15665 &{\bf 0.}03066 &{\bf 0.}15963 \\ 
     1  & 2230  &{\bf -0.}17802 &{\bf 0.}09517 &{\bf 0.}20186 \\ 
     1  & 9821  &{\bf -0.2}3282 &{\bf 0.2}3871 &{\bf 0.3}3345 \\ 
     1  &35741  &{\bf -0.2}1775 &{\bf 0.2}6485 &{\bf 0.34}287 \\ 
     1  &84139  &{\bf -0.2}2011 &{\bf 0.2}6071 &{\bf 0.34}120 \\ 
     1  &299690 &{\bf -0.21}847 &{\bf 0.2}6942 &{\bf 0.34}687 \\ 

\hline \hline
\multicolumn{5} {|l|}{JCMharmony 3D, second order elements (TM)} \\
\hline
n &N & $\Re (FC_0)$ & $\Im (FC_0)$ & $|FC_0|$ \\ 
     2  & 1884  &{\bf -0.2}6261 &{\bf 0.2}2353 &{\bf 0.34}486 \\ 
     2  &13672  &{\bf -0.2}2516 &{\bf 0.2}6312 &{\bf 0.34}631 \\ 
     2  &54398  &{\bf -0.21}984 &{\bf 0.2}6810 &{\bf 0.34}671 \\ 
     2  &99468  &{\bf -0.21}885 &{\bf 0.271}10 &{\bf 0.34}841 \\ 
     2  &376786  &{\bf -0.219}10 &{\bf 0.271}45 &{\bf 0.34}884 \\ 
\hline \hline
\multicolumn{5} {|l|}{JCMharmony 3D, third order elements (TM)} \\
\hline
n &N & $\Re (FC_0)$ & $\Im (FC_0)$ & $|FC_0|$ \\ 
     3  & 5652  &{\bf -0.2}2476 &{\bf 0.2}6408 &{\bf 0.34}678 \\ 
     3  &27264  &{\bf -0.21}862 &{\bf 0.271}99 &{\bf 0.34}896 \\ 
     3  &139659  &{\bf -0.219}71 &{\bf 0.271}99 &{\bf 0.349}64 \\ 
\hline \hline
\multicolumn{5} {|l|}{JCMharmony 3D, fourth order elements (TM)} \\
\hline
n &N & $\Re (FC_0)$ & $\Im (FC_0)$ & $|FC_0|$ \\ 
     4  &12608  &{\bf -0.21}726 &{\bf 0.27}308 &{\bf 0.34}896 \\ 
     4  &100036  &{\bf -0.219}02 &{\bf 0.271}54 &{\bf 0.34}886 \\ 
     4  &365384  &{\bf -0.219}31 &{\bf 0.2717}0 &{\bf 0.349}17 \\ 
\hline
\end{tabular}
\caption{Comparison of quasi-exact results (first row, see Reference~\cite{Burger2005bacus})
to results obtained using adaptive 3D FEM. 
The 3D results converge towards the quasi-exact results for increasing 
finite element degree $n$ and for increasing number of degrees of freedom $N$ (i.e., increasing 
grid refinement). 
Real  and imaginary parts and magnitudes  of the 
$0^{th}$ far field coefficients computed with 2D and 3D FEM are given in 
units [V/m], {\it cf}~\cite{Burger2005bacus}. Converged digits are marked in bold 
face.}
\label{bm_table_1}
\end{center}
\end{table}

\section{Conclusions}
\label{conclusions}
We have performed rigorous 3D FEM simulations of light transition through 
a large 3D photomask (size of the computational domain about 1000 cubic 
wavelengths). 
We have achieved results at high numerical accuracy 
which compare well to experimental 
findings using aerial imaging. 
We have checked the convergence for this 3D case and we have checked 
the convergence of the method for a simpler case, where a 
quasi-exact result is available. 
Our results show that rigorous 3D mask simulations can well 
be handled at high accuracy and relatively low computational cost.

\acknowledgements
We thank Arndt C. D{\"u}rr (AMTC Dresden) for the AIMS measurement.
The work for this paper was supported by the EFRE fund of the European Community and
by funding of the State Saxony of the Federal Republic of Germany (project number
10834). The authors are responsible for the content of the paper.

\bibliography{/home/numerik/bzfburge/texte/biblios/phcbibli,/home/numerik/bzfburge/texte/biblios/group,/home/numerik/bzfburge/texte/biblios/lithography}   
\bibliographystyle{spiebib}   

\end{document}